\documentclass[lettersize,journal]{IEEEtran}
\usepackage{amsmath,amsfonts}
\usepackage{algorithmic}
\usepackage{algorithm}
\usepackage{array}
\usepackage[caption=false,font=normalsize,labelfont=sf,textfont=sf]{subfig}
\usepackage{textcomp}
\usepackage{stfloats}
\usepackage{orcidlink}
\usepackage{url}
\usepackage{verbatim}
\usepackage{graphicx}
\usepackage{cite}
\usepackage{longtable}
\usepackage{booktabs}
\usepackage{hyperref}
\usepackage{changepage}
\hypersetup{
colorlinks=true,
linkcolor=blue,
citecolor=blue,
urlcolor=blue,
linktoc=all
}

\begin{document}

\title{Emotion Alignment: Discovering the Gap Between Social Media and Real-World Sentiments in Persian Tweets and Images}


\author{Sina Elahimanesh \textsuperscript{\orcidlink{0000-0001-7251-6661} \textasteriskcentered}, Mohammadali Mohammadkhani \textsuperscript{\orcidlink{0009-0008-1735-0062} \textasteriskcentered}, and Shohreh Kasaei \textsuperscript{\orcidlink{0000-0002-3831-0878}}
\thanks{S. Elahimanesh, M. Mohammadkhani, S. Kasaei are with the Department of Computer Engineering, Sharif University of Technology, Tehran, Iran. e-mails:\{sina.elahimanesh, mo.mohammadkhani, kasaei\}@sharif.edu.}
}



\maketitle

\begin{abstract}
In contemporary society, widespread social media usage is evident in people’s daily lives. Nevertheless, disparities in emotional expressions between the real world and online platforms can manifest. We comprehensively analyzed Persian community on X to explore this phenomenon. An innovative pipeline was designed to measure the similarity between emotions in the real world compared to social media. Accordingly, recent tweets and images of participants were gathered and analyzed using Transformers-based text and image sentiment analysis modules. Each participant's friends also provided insights into the their real-world emotions. A distance criterion was used to compare real-world feelings with virtual experiences. Our study encompassed N=105 participants, 393 friends who contributed their perspectives, over 8,300 collected tweets, and 2,000 media images. Results indicated a 28.67\% similarity between images and real-world emotions, while tweets exhibited a 75.88\% alignment with real-world feelings. Additionally, the statistical significance confirmed that the observed disparities in sentiment proportions.
\end{abstract}

\begin{IEEEkeywords}
Social Computing, Text Sentiment Analysis, Image Sentiment Analysis, Social Network, Social Media
\end{IEEEkeywords}

\renewcommand{\thefootnote}{\textasteriskcentered}
\footnotetext{These authors contributed equally.}

\section{Introduction}
\label{sec:intro}

Social networking platforms are now an integral part of daily life, particularly among youth, who frequently use these networks \cite{twenge2019trends}. A key reason for engagement is the expression of viewpoints and emotions through diverse textual or multimedia content on various platforms \cite{bala2014social}. The significant interaction with social media and responses to events have prompted numerous studies analyzing these reactions \cite{salgado2019news}. Advances in technology, especially smartphones and the Internet, have shifted daily communications to social networks \cite{srivastava2005mobile}, increasing reliance on these platforms. Social networks have adapted by releasing applications and messengers to enhance communication \cite{ortiz2023rise, appel2020future}, encouraging even greater engagement \cite{claussen2013effects}. This shift has led to increased time spent on social networks \cite{scott2017time}, making them valuable sources for identifying people’s behavior and their virtual characteristics.
However, people may not always express their true emotions and personality on social networks \cite{balfe2010disclosure}, or they may display different emotions to various audiences \cite{zheng2020self, vitak2012impact}. This behavior can stem from reasons such as shyness or insecurity about revealing true emotions \cite{fang2017coping}. Therefore, social networks cannot consistently serve as the sole source for detecting users’ personalities but can act as an auxiliary tool to help researchers explore different aspects of users’ personalities. 

One of the key aspects of personality is emotions or feelings \cite{montag2017primary}; thus, analyzing emotions offers insights into an important aspect of personality, reducing the problem to analyzing emotions. Emotions vary across environments and are influenced by context \cite{hess2020bidirectional}, with social influences significantly shaping emotional expression \cite{fischer2003social}. Individuals may express emotions differently across situations \cite{sherman2015independent}. By focusing on emotions across environments, this study proposes comparing emotions in real life and social media, addressing RQ \ref{RQ}. While some studies explore causal links between real-world activities and social media emotions \cite{mirlohi2021causal}, evidence on their alignment or gap is lacking. This study bridges this gap using a novel pipeline combining people's insights and AI-based modules, aiming to illuminate human emotional expression in virtual spaces and provide insights for researchers in the digital age. We came up with the following research question (RQ \ref{RQ}):

\begin{itemize}
    \item \textbf{RQ: Do people have the same emotions in social media compared to the real world?}\label{RQ} 
\end{itemize}

To address RQ \ref{RQ}, we proposed an innovative pipeline. It begins by collecting tweets and images from participants, analyzed via text and image sentiment modules to detect social media emotions. Simultaneously, real-life emotions are gathered from participants' friends. Statistical tests, particularly independent two-sample t-tests, evaluate whether differences in sentiment proportions between real life and social media are significant. Based on these findings, a distance criterion measures the disparities. Finally, results are presented on a website using charts and visual aids, ensuring transparency and addressing ethical and privacy concerns throughout the process.
The focus of the our analysis in this paper is on Persian-speaking users on the platform \textit{X}, selected for their large and active community \cite{zhou2010information}. To categorize emotional expressions on the platform and in real life, interviews were conducted with 10 active Persian-speaking users. Participants provided input on representative emotion categories, leading to five primary labels: \textit{Happiness}, \textit{Sadness}, \textit{Anger}, \textit{Neutral}, and \textit{Intense Emotions} (e.g., Love, Nostalgia, Surprise). These labels reflect a balance of two positive, two negative, and one neutral emotion, with suggestions endorsed by fewer than five participants excluded for focus.

The contributions of our work are threefold: (1) a novel dataset was collected from Persian X, comprising over 3000 records (tweets and images) labeled into 5 sentiments (accessible in [\href{https://www.kaggle.com/datasets/mohammadalimkh/persian-twitter-dataset-sentiment-analysis/data}{Kaggle}]). Using this dataset, a Persian Transformers-based text sentiment analysis module classified tweet sentiments into 5 predefined classes with an accuracy of 74.08\%; (2) a novel pipeline was defined to address RQ \ref{RQ}, integrating AI-based modules with human perspectives; (3) an experiment was conducted with N=105 Persian-speaking participants and 393 individuals as their friends. Findings showed a 75.88\% similarity between participants' real-life emotions and the sentiments expressed in their tweets, but a lower similarity (28.67\%) between their real-world emotions and the sentiments conveyed by their images on X. Following the experiment, a comprehensive qualitative and quantitative analysis evaluated the pipeline's efficiency and participant satisfaction, with about 93\% of users agreeing that the analysis accurately reflected their feelings.

\section{Related Work}

Our work is motivated by the literature on finding the relation between social network emotions and real-world emotions. First, cognitive studies conducted around this subject are mentioned, as well as their conclusions and results. After that, we will go over research projects involving the utilization of artificial intelligence in the process of emotion detection on social networks in general. Then, we focus on studies regarding X and review the methods and approaches proposed to perform sentiment analysis on tweets.

\subsection{People's Emotions in Social Network and Its Relation to the Real-World Emotions}
\label{ssec:four-limitations}

With the emergence of Social Network Analysis (SNA) \cite{clifton2017introduction}, researchers began to investigate the intricate relationships between individuals' real-world behaviors and their activities on social media platforms \cite{staiano2012friends, selfhout2010emerging, wehrli2008personality}. Early research efforts aimed to uncover how online behavior could reflect or even influence offline characteristics and life circumstances \cite{amichai2010social}. These explorations extended into various domains of everyday life, including the dynamics of romantic relationships \cite{neyer2004personality}, leadership and management capabilities \cite{emery2013leadership}, and mental health conditions such as depression and anxiety \cite{keles2020systematic}. Observing consistent behavioral patterns across both digital and physical contexts motivated further studies into the feasibility of inferring psychological traits directly from social media data. As a result, researchers began developing models and methodologies for automatically extracting users’ personality traits based on their online interactions and digital footprints \cite{qiu2012you, celli2011mining, tai2016systematical}. This line of inquiry has opened new avenues for understanding human behavior at scale, enabling personalized services, mental health assessments, and social computing applications.

\subsection{Contribution of AI-based Methods in Social Network Analysis}

Before the advent of deep learning, researchers have proposed methods to predict emotions on social networks \cite{golbeck2011predicting}. These methods include feature extraction \cite{celli2012adaptive}, XGBoost classifiers \cite{khan2020personality, kunte2020personality}, SVM \cite{hasan2018machine, gautam2014sentiment}, and Naive Bayes \cite{alam2013personality} to detect emotions on social media. As time passed, new techniques and models were built based on Neural Networks \cite{abbas2021social, xue2018deep, mehta2020recent}, Contextual Language Embeddings \cite{kn2021latent}, Convolutional Neural Networks (for image analysis) \cite{segalin2017social}, and the transformer architecture \cite{vaswani2017attention} to address this task. Among these advancements, transformer-based models played a vital role in addressing the task of emotion detection and sentiment analysis \cite{naseem2020transformer, arijanto2021personality}, or even in computer vision tasks such as facial expression \cite{zhang2022transformer}. Later in 2018, with the development of GPT-1 as the first Large Language Model (LLM), a new era started for natural language processing tasks (including sentiment analysis) \cite{zhang2023sentiment, gupta2024comprehensive, zhan2024optimization, appaji2024leveraging}, as these tasks became achievable with high accuracy through the use of simple prompts \cite{briganti2024chatgpt}. Alongside monolingual text sentiment analysis \cite{ipa2024bdsentillm}, LLMs have been utilized to perform the multilingual sentiment analysis task \cite{miah2024multimodal} as well.
These technologies helped researchers explore different issues that require sentiment analysis on social media, such as classifying travel behaviors \cite{cui2018travel}, recognizing hate speech in Arab communities \cite{aljarah2021intelligent}, classifying age groups in social networks \cite{guimaraes2017age}, and other applications \cite{yen2021detecting, bari2020approach}.

\subsection{Prior Analysis Studies on X}

This section focuses exclusively on X, excluding studies on other social networking platforms. Early research on X, similar to other social networks, relied on classical machine learning techniques such as SVM \cite{hasan2018machine, gautam2014sentiment}, TF-IDF \cite{kumar2019personality}, Naive Bayes \cite{sailunaz2019emotion}, and KNN \cite{pratama2015personality}. Other approaches like LSTM \cite{karami2023transformer}, clustering \cite{yousefi2023early} and CNN \cite{dastgheib2020application} have also been applied for analyzing X. But later on, with the emerge of transformer models and LLMs, several research works were conducted on X based on these concepts \cite{naseem2020transformer, shi2023bert, arijanto2021personality, jaradat2024multitask, gambini2024evaluating}. These advancement in sentiment analysis has led to a lot of applications, such as hate speech recognition \cite{ayo2020machine}, discussions about COVID-19 \cite{xue2020twitter, muller2023covid}, and sarcasm detection \cite{sarsam2020sarcasm}.

Additionally, these methods have also been leveraged to perform sentiment analysis on X within the Persian-speaking community \cite{fatehimbti, heydari2021deep}. Building on this foundation, various downstream applications have been developed, including emoji prediction \cite{tavan2020persian}, which aims to enhance user interaction modeling, as well as fake news detection \cite{sadr2020predictive, sadr2021use}, which addresses the growing concern over misinformation on social platforms. Other significant applications include rumor detection \cite{mahmoodabad2018persian}, which is critical for maintaining information integrity, and a range of additional use cases related to content mapping and user behavior analysis on Persian Twitter \cite{khazraee2019mapping, nezhad2022twitter}.

\section{Proposed Method}

In this section, essential prerequisites regarding the analysis are initially discussed, and then an overview of the method is presented. Subsequently, the next subsections discuss various stages within the pipeline in detail.

\begin{figure*}
  \includegraphics[width=\textwidth]{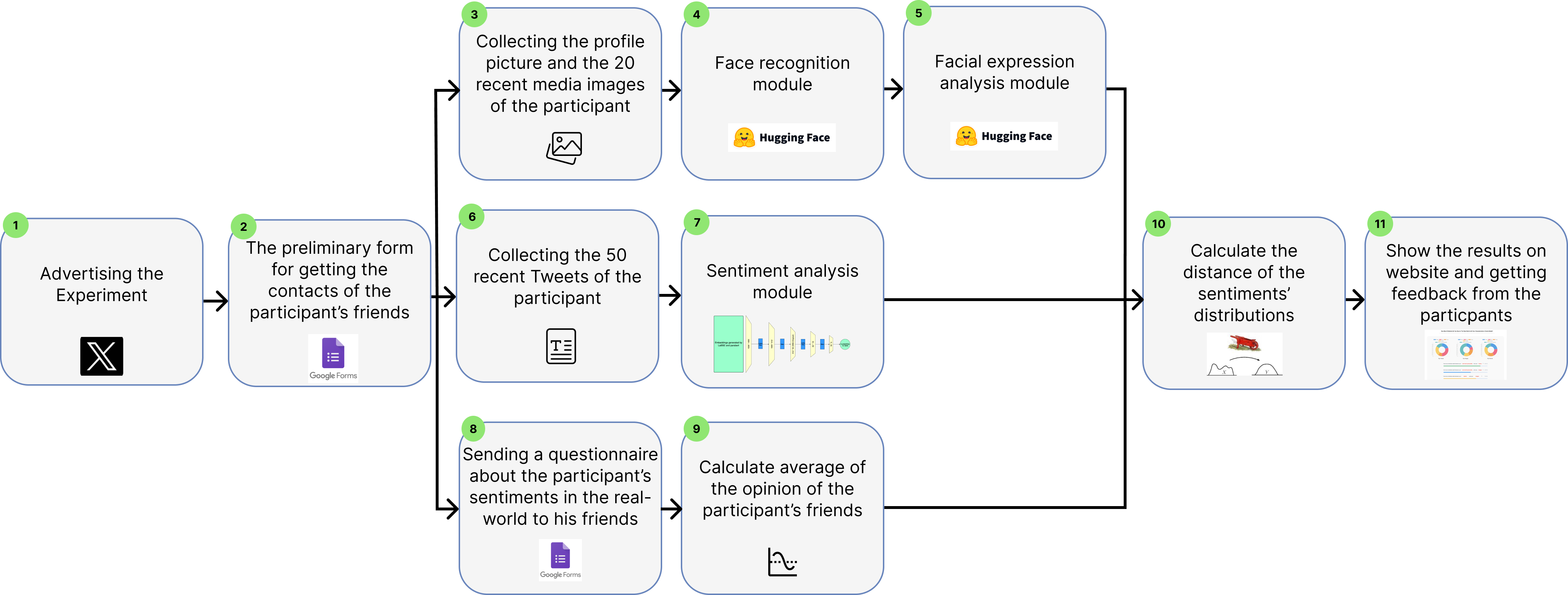}
  \caption{Overview of the experimental pipeline for analyzing participants' online and offline emotional perceptions. The study is initiated by advertising the experiment on social media platforms such as X platform (Step 1), followed by collecting contact information of participants’ friends through a preliminary Google Form (Step 2). The system then gathers visual and textual data, including the participant’s profile picture and 20 recent media images (Step 3), and 50 recent tweets (Step 6). Face recognition (Step 4) and facial expression analysis (Step 5) modules, powered by Hugging Face models, are applied to the images. In parallel, a sentiment analysis module processes the textual content of tweets (Step 7). A separate questionnaire is sent to the participant’s friends to capture their perception of the participant’s real-world sentiments (Step 8), which is then averaged (Step 9). The system calculates the distance between the sentiment distributions derived from online data and friends’ responses (Step 10). Finally, the results are visualized on a dedicated website and participants are invited to provide feedback (Step 11).}
  \label{fig:pipeline}
\end{figure*}

\subsection{An Overview of The Proposed Method}
Our proposed pipeline consists of 11 main stages. As shown in Figure \ref{fig:pipeline}, first, the experiment is being advertised. We use X as the main channel of the advertisement. In such a case, the advertisement is posted on our X accounts, and people are encouraged to participate in the experiment. The advertisement tweet was reposted many times, and the main participants were collected using this approach. 

In Stage 2, each participant is sent a Google form to gather some information, such as the participant's age and name. Before sending the forms, the experiment process is explained in full detail to the potential participants, and their consent is gathered. The objective of the form is to collect the primary information of the participants and the contact information of their friends. It is worth noting that in this form, it is clearly mentioned that participants can introduce their close friends, family members, partners, or anybody who knows them sufficiently in the real world and not in a remote condition. Also, a participant who fills out the mentioned form consents to attend the experiment and consequently accepts that their tweets and photos will be crawled for the experiment.

As shown in Stages 3 and 6, 50 tweets are collected from the user’s recent tweets as well as the participant's profile image and 20 recent images published in their account on X. To ensure the relevance and consistency of the data, participants were filtered to include only those with at least 50 recent tweets and a minimum of 20 images published in their account. The tweets and images collected are the most recent ones based on the number limit.
Simultaneously with the crawling stage, in Stage 8, another Google Form is sent to the participant's friends to ask them some quantitative straightforward questions about the emotions of the participant in the real world. Then, in Stage 9, an average of the gathered data for each participant is calculated as the representative of their sentiments in the real world.

In Stages 4 and 5, the images go through the face recognition module, and then the output is used as input for the facial expression analysis module to find the sentiment of the images, which is classified into 5 predefined classes. Meanwhile, in Stage 7, the crawled tweets are all given to our Persian sentiment analysis module to detect and classify participants' emotions into the same 5 classes.

Following the extraction of three vectors with 5 values regarding participants' feelings in their real world, tweets, and images, a distance criterion is employed to identify the level of similarity between them, as shown in Stage 10.
Finally, at the last stage, participants are informed of the conducted analysis via a comprehensive website with various visual components. Afterward, they are asked to fill out a Google Form, encompassing qualitative and quantitative questions, to give feedback about the experiment. 
In the rest of this section, each module is discussed in more depth and with enough detail.

\subsection{Real-World Sentiments Extraction}
This stage extracts real-world sentiments by surveying individuals close to the participant rather than the participant themselves, avoiding potential self-bias \cite{van2008faking, ong2018happier}. Participants select 3–5 close contacts to provide insights into their emotions. These friends receive a form with metadata and five questions, each targeting one of the five sentiment categories identified in the analysis. The questions focus on recent behaviors, with responses rated on a 1–10 scale, ensuring alignment with recent tweets and images. The collected data generates a sentiment distribution, normalized into probabilities. This approach leverages friends' deeper understanding of the participant's emotional tendencies. An example question posed to the participant’s friends is as follows, with \textit{Happy} substituted for other emotions as needed: \textbf{How frequently or to what extent do you believe your friend has been Happy in the real world recently? (The response is a numerical value within the range of 1 to 10.)}

\begin{figure*}
  \includegraphics[width=\textwidth]{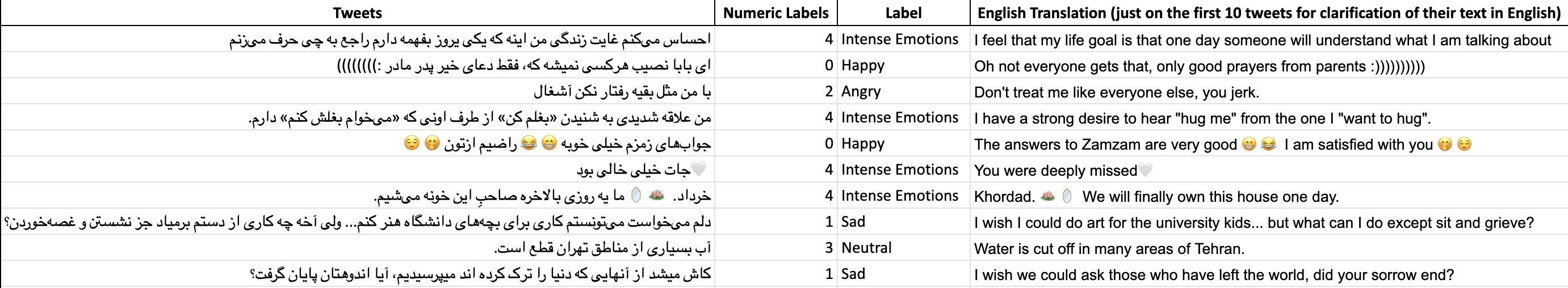}
  \caption{A screenshot of 10 tweets from the gathered dataset is shown here.}
  \label{fig:sampletweets}
\end{figure*}

\subsection{Collected Dataset}

\textit{\textbf{Dataset Gathering:}}
Owing to the absence of a suitable Persian X dataset, we decided to create our own dataset by collecting data from Persian-speaking users on X. 
We used the Snscrape \footnote{https://github.com/JustAnotherArchivist/snscrape} package to crawl tweets. Snscrape is a Python package that is designed to legally crawl data from social networking applications, including X, without violating any rules regarding privacy and policy. Using this package, over 3300 raw tweets were collected from different users. After the raw data were collected, we anonymized the data to ensure that users' privacy and confidentiality are not affected. After that, the first two authors independently annotated all records in the dataset using a predefined set of five distinct labels. In the rare cases where labeling disagreements arose, affecting only 0.1\% of the tweets, conflicts were resolved through discussion and consensus.

\textit{\textbf{Data Preprocessing:}}
We used a python package called \textit{Hazm}, which is a module used for normalizing Persian texts. This process helps the model make a better decision since all of the tweets in our dataset are preprocessed and normalized according to the standards of the Persian language. Alongside Hazm, we have also used Pandas library functions to perform EDA on our dataset. These functions help us detect duplicate or empty tweets if they exist in the dataset.
Finally, after these stages of anonymization, normalization, and labeling, the dataset is ready to be used and fed to text analysis models. The dataset is published and accessible through this link \footnote{https://www.kaggle.com/datasets/mohammadalimkh/persian-twitter-dataset-sentiment-analysis/data}. Also, 10 sample tweets from the dataset are shown in Figure \ref{fig:sampletweets}. Note that the English translation is added for these 10 tweets so that their meanings are clear.

\begin{figure*}
  \includegraphics[width=\textwidth]{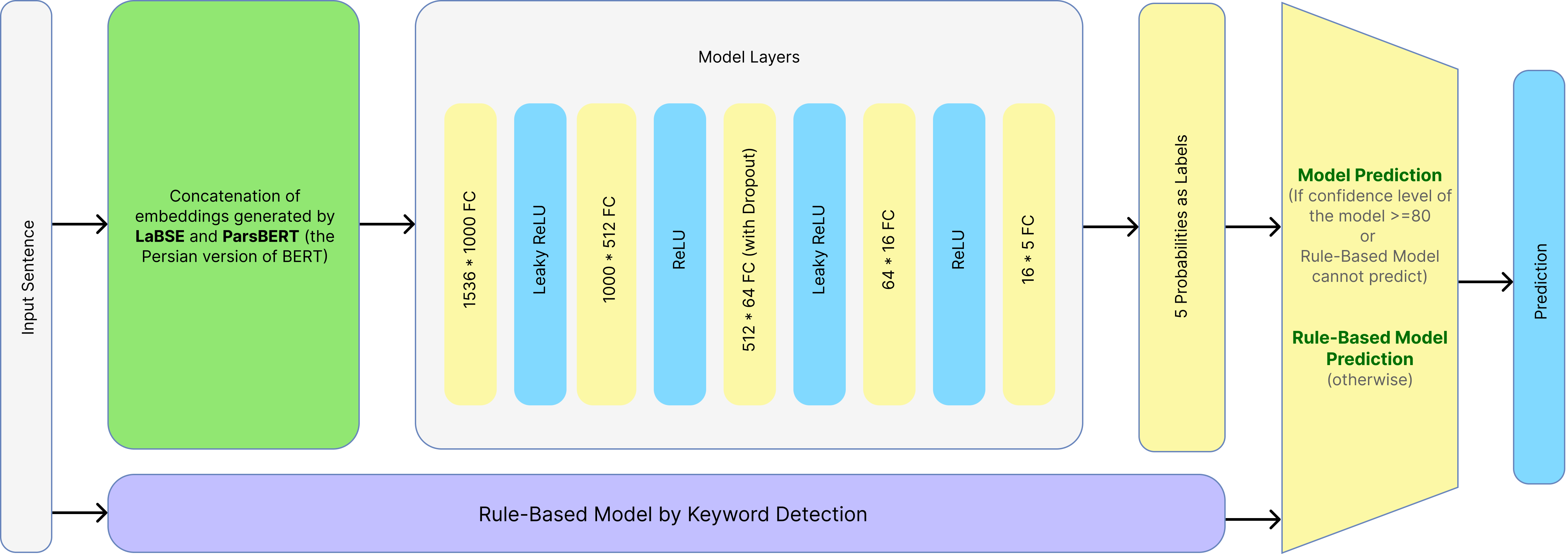}
  \caption{Architecture of the final hybrid sentiment classification model. The model takes a Persian input sentence, generates two sets of contextual embeddings using ParsBERT and LaBSE models, and concatenates the generated embeddings into a single vector. The final embedding vector is passed through a series of fully connected layers and non-linear activation functions (e.g., ReLU and Leaky ReLU) to produce a probability distribution over five sentiment classes. In parallel, a rule-based model based on keyword detection attempts to predict sentiment. The final output is selected based on a decision rule: if the neural model’s confidence is $\geq 80$ or the rule-based system cannot make a prediction, the model’s output is used; otherwise, the rule-based result is chosen.}
  \label{fig:arch}
\end{figure*}

\subsection{Social Media Sentiments Extraction}

\textit{\textbf{Text Sentiment Analysis Pipeline:}}
    To make tweets understandable for models, embeddings (which are vectors that represent the meaning of sentences) are generated using two models. The first model, ParsBERT \cite{ParsBERT}, is a pre-trained Persian version of BERT developed by HooshvareLab, which produces a 768-dimensional vector for each input sentence or paragraph. The second model, LaBSE \cite{feng2020language}, is a multilingual embedding model that maps texts from multiple languages into a shared vector space, also producing a 768-dimensional vector. These embeddings are concatenated to form a unified 1536-dimensional vector for each tweet, preparing it for entering the classifier model.
    With each tweet represented as a 1536-dimensional vector, a classifier model is required to output five probabilities corresponding to predefined labels. While various classifiers like SVM, Naïve Bayes, and Random Forest are available, a fully connected neural network was chosen due to its superior performance, as displayed in Table \ref{text:accuracies}. This neural network maps the 1536-dimensional input vectors to five probabilities, illustrated in Figure \ref{fig:arch}.
   To improve our method's performance, a set of predefined keywords is created for each label. Each of these keywords is mostly used and repeated in a unique sentiment, leading to better differentiation between sentiments. After processing, the classifier assigns each tweet a predicted label and a \textbf{confidence score} generated by PyTorch. If the confidence score is 80\% or higher, the predicted label is directly output. Otherwise, the system checks for keywords in the tweet. If a relevant keyword is found, the corresponding label is assigned. If no keyword is present, the system defaults to the classifier's prediction. The decision process is illustrated in Figure \ref{fig:arch}. The proposed model achieves an accuracy of 74.08\%, with an F1 score of 72.02\%, precision of 72.03\%, and recall of 75.99\%. 

\textit{\textbf{Image Sentiment Analyses Pipeline:}}
The image processing pipeline includes two main components. The first is a face recognition model from Hugging Face \footnote{https://huggingface.co/spaces/ParisNeo/FaceRecognition}, which uses a participant's profile image to identify and locate their face in collected images. This enables the extraction of facial regions for further analysis. The second component is a facial expression analysis model, also from Hugging Face \footnote{https://huggingface.co/spaces/schibsted/Facial\_Recognition\_with\_Sentiment\_Detector}, which detects sentiments in the extracted facial images using a pre-trained VGG19 model trained on the FER2013 dataset. The model's seven basic emotions are mapped to five categories by merging similar emotions (\textit{Sad} with \textit{Fear}, and \textit{Angry} with \textit{Disgust}). The final output is an averaged probability vector of the participant’s sentiments across all analyzed images.

\begin{table*}[!h]
\centering
  \caption{The evaluation metrics for various models on test data are indicated below.}
\begin{tabular}{ccccc}
 \toprule
\textbf{Classification Model} & \textbf{Accuracy} & \textbf{F1 Score} & \textbf{Precision} & \textbf{Recall}  \\ \hline
\textbf{Baseline}        & 20.00\% & 20.00\% & 20.00\% & 20.00\%  \\ \hline
\textbf{Random Forest}        & 48.00\% & 48.90\% & 51.04\% & 47.06\%      \\ \hline
\textbf{Transformers Encoder-Decoder}        & 51.02\%  & 50.53\% & 53.12\% & 48.22\%    \\ \hline
\textbf{Fine-Tuned Sentiment Analyzer}        & 51.49\%  & 48.88\% & 49.79\% & 50.92\%   \\ \hline
\textbf{Our Proposed Method}        & 74.08\% & 72.02\% & 72.03\% & 75.99\%  \\    \bottomrule
\end{tabular}
\label{text:accuracies}
\end{table*}

\subsection{Sentiment Comparison Metric}

The result from previous stages consists of three discrete distributions, each containing 5 values corresponding to different sentiment classes. These distributions represent the user’s tweets, user’s images, and their real-word emotions. To measure the similarity or distance between these distributions, we calculate the distances between each pair.
For an effective similarity measure, the chosen distance criterion must satisfy two main requirements: (1) it should be sensitive to both large and small differences between distributions, detecting even minor variations; and (2) it should be bounded, allowing similarity to be reported on a scale from 0 to 100. Although similarity can be derived by subtracting the distance value from 100, the selected criterion needs to fulfill these conditions. We considered three distance metrics and ultimately chose \textit{Earth Mover's Distance} for its suitability, with the formula of $D_{EMD}(P, Q) = \sum_{i} |P(i) - Q(i)|$, where \( P \) and \( Q \) are the discrete distributions.
Details of the two other metrics considered but not selected are in Appendix \ref{appendix:distance_metrics}. Earth Mover’s Distance (EMD) \cite{andoni2008earth} was chosen for its sensitivity to small changes, boundedness, and ability to minimize transportation cost. In contrast, KL-Divergence is less effective due to its inability to handle small variations and lack of boundedness. Jensen-Shannon Divergence is bounded but fails to capture small differences sufficiently.

\subsection{Feedback Collection Website}

\begin{figure*}
  \includegraphics[width=\textwidth]{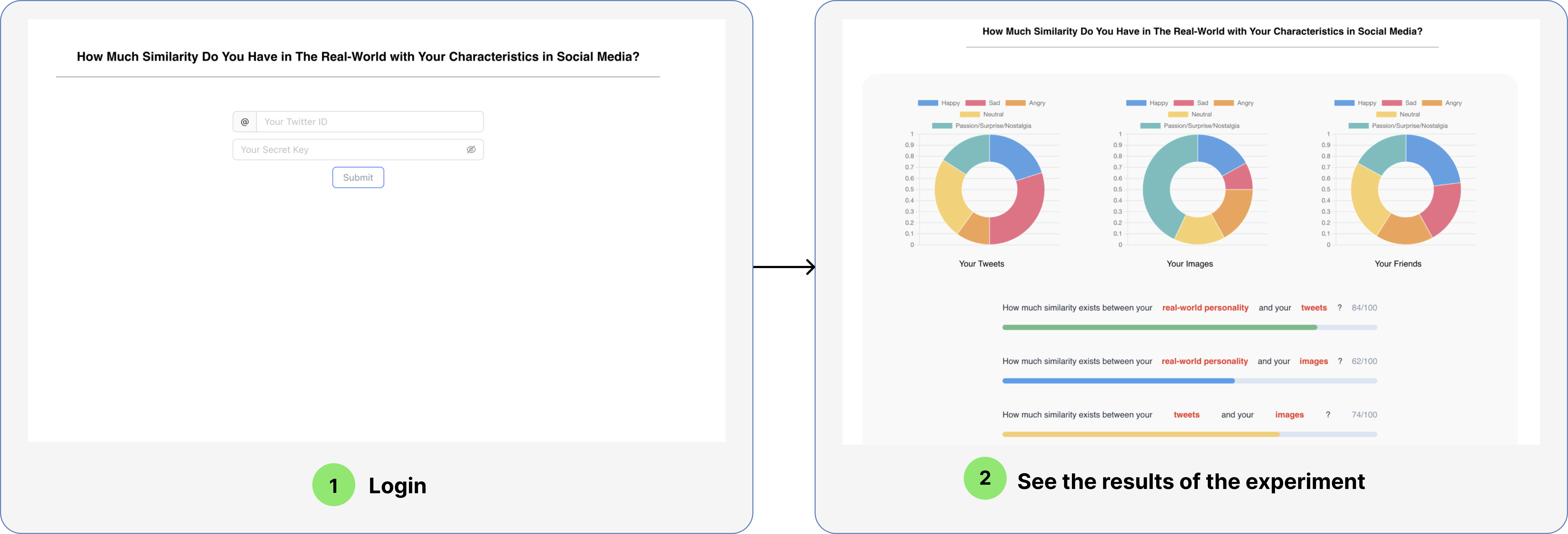}
  \caption{Overview of the website for the experimental process for assessing the our results in a post-survey study. The left panel (Step 1: Login) depicts the user authentication interface, where participants input their ID on X platform and a secret key to grant access to their social media data (which were given to them as a password). The right panel (Step 2: See the results of the experiment) displays the outcome of the analysis through visualizations comparing emotional expressions and personality features across three dimensions: participants’ tweets, uploaded images, and their friends’ content. The results include donut charts illustrating the distribution of emotional tones (e.g., happy, sad, neutral, angry) and bar charts that quantify the similarity between the user’s real-world personality and their social media expressions. Each similarity score is presented on a scale from 0 to 100, highlighting how consistently individuals portray themselves across different social media modalities.}
  \label{fig:website}
\end{figure*}

To design a website for users to review their results, we need to define a backend stack and a frontend stack. By integrating these two stacks, the user interface interacts with the database and backend systems to access users' data. The choice for backend technology is Django, which is a framework built based on Python. It is one of the most powerful technologies for handling the logic and the backend side of websites. On the other side, React is used to design the user interface and connect it to the backend. React is an efficient framework developed by Facebook to handle the frontend side of websites. In our design, an important requirement is considered, which is the responsiveness of website design, meaning that the design is not affected and disturbed by switching between different types of devices for viewing the website. After the completion of design and development, the code for both the front end and back end is deployed onto a server that has been previously acquired. Then, the address of the website is sent to participants so that they can observe their results.
The flow of the website is shown in Figure \ref{fig:website}. First, participants enter their X ID and their exclusive credentials, which were previously handed out to them. If the inputted information is correct, participants are navigated to the main screen, where their results are ready to be demonstrated. The findings are presented in three separate charts, each chart representing information obtained from tweets, images, and real-world opinions. Below the charts, users can observe three distinct similarity scores on a scale of 0 to 100, derived from the pairwise distances computed between tweets, images, and friends' assessments. These numbers report to users how much similarity exists between their emotions on X and their feelings in the real world.

\subsection{Ethics and Privacy Concerns}

We have addressed all ethical considerations throughout this project. Participants were fully informed about the purpose and procedures of the experiment, including the collection and publication of their tweet data. We obtained explicit consent from each participant for their tweets to be anonymized and used in the study. This consent was granted through their voluntary participation in the experiment. Additionally, participants were aware that their anonymized data, including sentiment distributions and distribution distances, would be published as part of our research. We also ensured that consent was obtained for the use of their friends' data, which was collected for supplementary analysis purposes. All participants participated voluntarily and received no incentives.

We have also paid enough attention to privacy by providing each user with their password to see their experiment results. Also, the thorough anonymization of the gathered tweets was ensured, and additional measures were taken to safeguard the identities of the individuals who authored them. In this project, another facet of privacy has been tackled: participants will remain unaware of which specific designated friend has provided reviews regarding their real-life emotions. This way, the privacy of participants' friends was protected. At the end of the experiment, each participant was provided with a form to express their sincere comments and feelings about every detail and stage of the experiment. All of the comments received were analyzed, and they contained no ethical issues. Everyone was satisfied with the procedure of the experiment and the way the ethical and privacy-related concerns were addressed. 

\section{Experimental Evaluation and Results}

The experiment involved N=105 Persian-speaking participants aged 17 to 33 years (average age = 21.84, STD = 2.13), including 63 males and 42 females. For each participant, 50 recent tweets (including their original tweets, replies, and quote tweets) and 20 profile/media images were collected. Tweets were analyzed using a text sentiment analysis module to calculate sentiment distributions as percentages. A Google form was also sent to 3–5 friends per participant (average: 3.74, totaling 393 people), all aged over 16.
Table \ref{tab:sentiments} presents the average sentiment distributions for participants' real-world emotions, social media images, and tweets. Participants appeared happiest or most passionate in photos, while tweets showed less happiness. Real-world sentiments were dominated by happiness, with anger being least common. In tweets, neutral sentiment was most frequent, while anger was least. For images, happiness was most common, and neutral was least. Detailed sentiment distributions and calculated distances are provided in Appendix \ref{appendix:sentiment_analysis}.

\begin{table*}[!h]
\centering
\caption{The results of the experiment, done on N=105 participants, along with the mean of each sentiment reported. We find nearly equally distributed emotions in the real world and the tweets; however, in the profile and media images, the majority of the images were classified in the two positive categories of happiness and intense emotions (Surprise, Love, and Nostalgia).}
\begin{tabular}{cccc}
\toprule
\textbf{Sentiment} & \textbf{Mean in the Real World}        & \textbf{Mean in Tweets' Text} & \textbf{Mean in the Profile and Media Images} \\ \hline
\textbf{Happiness}        & 24.37\%      & 16.79\%                                                               & 48.40\%          \\ \hline
\textbf{Sadness}        & 18.11\% & 22.65\%                                                             & 8.36\%        \\ \hline
\textbf{Neutral}        & 20.14\% & 24.36\%                                                         & 3.94\%      \\ \hline
\textbf{Anger}        & 13.67\% & 15.76\%                                                             & 4.50\%        \\ \hline
\textbf{Intense Emotions}        & 22.78\% & 20.46\%                                                             & 34.84\%         \\ \hline
\textbf{} & \textbf{STD in the Real World}        & \textbf{STD in Tweets' Text} & \textbf{STD in the Profile and Media Images} \\ \hline
\textbf{Happiness}        & 6.79\%      & 8.75\%                                                               & 21.65\%          \\ \hline
\textbf{Sadness}        & 6.14\% & 9.96\%                                                             & 8.44\%        \\ \hline
\textbf{Neutral}        & 6.02\% & 9.87\%                                                         & 5.68\%      \\ \hline
\textbf{Anger}        & 4.84\% & 7.85\%                                                             & 4.42\%        \\ \hline
\textbf{Intense Emotions}        & 5.35\% & 8.27\%                                                             & 18.57\%         \\ \bottomrule
\end{tabular}
\label{tab:sentiments}
\end{table*}

\subsection{Sentiment Similarities}
Table \ref{tab:sentiments} presents key findings. The average similarity between participants' tweet sentiments and real-world emotions was 75.88\%, indicating that tweets reliably represent real-world emotions. In contrast, the similarity between image contents and real-world emotions was significantly lower at 28.67\%, highlighting that people often display happiness or positivity in photos while their real-world emotions are more evenly distributed across sentiments.
The analysis confirms that text content serves as a reliable indicator of real-world emotions, whereas images do not. Additionally, the average similarity between sentiments in participants' images and tweets was only 38.02\%, further emphasizing the disconnect between these mediums.

\begin{table*}[!h]
\centering
\caption{The average similarity of the sentiments of the participants in different configurations by considering the complement of the Earth Mover's distance metric as the proposed similarity metric.}
\begin{tabular}{ccc}
\toprule
\textbf{Sentiments' Similarity (Earth Mover's Distance)} & \textbf{Mean} & \textbf{STD} \\ \hline
\text{Between the text of participants' tweets and their real-world}        & 75.88\%  & 14.69\%          \\ \hline
\text{Between the profile image and media images of participants and their real-world}        & 28.67\%    & 30.01\%       \\ \hline     
\text{Between the profile image and media images of participants and their tweets' text}        & 38.02\%  & 29.78\%          \\ \bottomrule     
\end{tabular}
\label{tab:sentiments}
\end{table*}

\subsection{Statistical Analysis on Sentiment Similarities}

To further validate the observed differences in sentiment distributions across various groups (images, tweets, and friends' perceptions), we conducted statistical tests on the data available in the appendix \ref{appendix:sentiment_analysis} to determine whether these differences are statistically significant. We employed the independent two-sample t-test to compare the means of sentiment proportions between the groups. The t-test assesses whether the means of two independent groups differ significantly, providing a t-statistic and corresponding p-value.

For each sentiment category, we performed pairwise comparisons between the sentiment proportions derived from participants' images, tweets, and friends' perceptions. The null hypothesis for each test was that there was no significant difference in the mean sentiment proportions between the two groups being compared.
The results of these t-tests are summarized in Table \ref{tab:statistical_tests}. The table provides the t-statistic and p-value for each pairwise comparison, as well as an indication of whether the difference is statistically significant at a typical alpha level of 0.05.

\begin{table*}[!h]
\centering
\caption{Results of the independent two-sample t-test comparing sentiment proportions between different groups.}
\begin{tabular}{cccccc}
\toprule
\textbf{Sentiment} & \textbf{Group 1} & \textbf{Group 2} & \textbf{T-statistic} & \textbf{P-value} & \textbf{Significance} \\ \hline
\textbf{Happy} & Images & Tweets & 13.85 & $2.43 \times 10^{-25}$ & Significant \\ \hline
\textbf{Happy} & Images & Friends & 10.87 & $7.74 \times 10^{-19}$ & Significant \\ \hline
\textbf{Happy} & Tweets & Friends & -7.16 & $1.19 \times 10^{-10}$ & Significant \\ \hline
\textbf{Sad} & Images & Tweets & -10.70 & $1.84 \times 10^{-18}$ & Significant \\ \hline
\textbf{Sad} & Images & Friends & -9.61 & $5.08 \times 10^{-16}$ & Significant \\ \hline
\textbf{Sad} & Tweets & Friends & 4.83 & $4.80 \times 10^{-06}$ & Significant \\ \hline
\textbf{Angry} & Images & Tweets & -13.61 & $7.77 \times 10^{-25}$ & Significant \\ \hline
\textbf{Angry} & Images & Friends & -14.24 & $3.55 \times 10^{-26}$ & Significant \\ \hline
\textbf{Angry} & Tweets & Friends & 2.71 & 0.0078 & Significant \\ \hline
\textbf{Neutral} & Images & Tweets & -17.64 & $4.97 \times 10^{-33}$ & Significant \\ \hline
\textbf{Neutral} & Images & Friends & -19.95 & $2.45 \times 10^{-37}$ & Significant \\ \hline
\textbf{Neutral} & Tweets & Friends & 4.00 & 0.00012 & Significant \\ \hline
\textbf{Intense Emotions} & Images & Tweets & 7.32 & $5.48 \times 10^{-11}$ & Significant \\ \hline
\textbf{Intense Emotions} & Images & Friends & 6.22 & $1.04 \times 10^{-08}$ & Significant \\ \hline
\textbf{Intense Emotions} & Tweets & Friends & -2.69 & 0.0083 & Significant \\ \bottomrule
\end{tabular}
\label{tab:statistical_tests}
\end{table*}

The results indicate that all pairwise comparisons across the three groups (images, tweets, and friends' perceptions) show statistically significant differences in sentiment proportions. The extremely low p-values across all comparisons suggest that the observed differences are not due to random chance. This reinforces the earlier observations that the sentiments expressed in images, tweets, and perceived by friends differ significantly.
These findings align with the hypothesis that people may present different emotional expressions in different mediums (e.g., images vs. text) and that external perceptions (from friends) also vary, potentially reflecting differences in how people choose to represent themselves or how they are perceived by others.

\subsection{Qualitative User Study}

Along with the website credentials, participants received a Google Form designed to assess the accuracy and efficiency of the analysis, featuring both quantitative and qualitative questions. One key quantitative question asked participants to rate, on a scale from 1 to 10, how accurately they believed the experiment analyzed their emotions and sentiments. The results showed that approximately 93\% of participants considered the analysis accurate and justified, suggesting a high level of reliability in the experiment’s outcomes. In addition to this, the qualitative feedback provided by participants offered valuable insights, which were thematically categorized into 10 clusters, as detailed in Table \ref{tab:qualitative}, with key conclusions summarized in as below:

\begin{itemize}
    \item Considering $Theme_1$, $Theme_2$, $Theme_6$, and $Theme_9$, participants expressed general satisfaction with the user experience, accuracy of the analysis, and the comprehensiveness of the questionnaires, evident as the absence of any negative comments regarding these factors. Also, the experiment was well-received, with users appreciating the awareness it brought to their emotions on social media.
    \item A primary area for improvement identified was the need to generalize the experiment by including a more diverse range of tweets and images based on $Theme_3$. Participants suggested implementing a time limitation for data crawling, limiting it to the date of the content rather than a fixed quantity, as highlighted in $Theme_5$.
    \item Participants in $Theme_4$ recommended extending the experiment to encompass a variety of platforms. Comparing and consolidating analyses from different social media platforms could potentially offer a more comprehensive understanding of participants' characteristics.
    \item Participants in $Theme_{10}$ highlighted the importance of personalization in the analysis, especially concerning participants' emotion types. Suggestions included tailoring pipelines for introverted versus extroverted individuals, even considering the possibility of involving introverted individuals directly rather than relying solely on input from their friends.
    \item Overall, participants expressed satisfaction with the privacy issues undertaken throughout the experiment. They appreciated the ethical considerations and transparency in all processes, which contributed to their awareness of the details of the methods employed, evident in $Theme_7$ and $Theme_8$.
\end{itemize}

The conducted qualitative analysis serves as a crucial tool for evaluating the effectiveness of the proposed pipeline. By gathering insights on participants' satisfaction, identified challenges, and suggested improvements, this analysis not only validates the current study but also provides valuable guidance for future research studies,
and it can define their potential direction.

\begin{table*}[!h]
\centering
\caption{Qualitative metrics of the evaluation.}
\begin{tabular}{ccc}
\toprule
\textbf{Theme Class} & \textbf{Overall Theme} & \textbf{Representative Sample}  \\ \midrule

\textbf{$Theme_1$} & \text{Fine User Experience} 
& \text{The user interface was elegant and very easy} \\ 
\textbf{} & \text{} 
& \text{to understand.} \\ 
\midrule

\textbf{$Theme_2$} & \text{Accurate Results} 
& \text{It was pretty accurate and close to the reality.} \\ 
\midrule

\textbf{$Theme_3$} & \text{Increasing Number of} 
& \text{I think maybe considering more tweets or images} \\ 
\textbf{} & \text{Analyzed Tweets and Images} 
& \text{can improve the system.} \\ 
\midrule

\textbf{$Theme_4$} & \text{Extending the Experiment to} 
& \text{It would be great to generalize it to various social} \\
\textbf{} & \text{Various Platforms} 
& \text{media platforms to have a more accurate understanding} \\
\textbf{} & \text{} 
& \text{of the user.} \\
\midrule

\textbf{$Theme_5$} & \text{Time-Boxed Analysis} 
& \text{I suppose that it's better to analyze tweets and images} \\
\textbf{} & \text{} 
& \text{based on their time, not a fixed number of them. For} \\
\textbf{} & \text{} 
& \text{example, they can be limited to recent 5 months.} \\
\midrule

\textbf{$Theme_6$} & \text{Comprehensive Questionnaires} 
& \text{The experiment was comprehensive with efficient} \\
\textbf{} & \text{} 
& \text{questionnaires, which did not have any extra questions.} \\
\midrule

\textbf{$Theme_7$} & \text{Transparent Procedures} 
& \text{It was pretty good that all procedures of the experiment} \\
\textbf{} & \text{} 
& \text{with details were transparent, and clearly explained in} \\
\textbf{} & \text{} 
& \text{getting my consent of participation.} \\
\midrule

\textbf{$Theme_8$} & \text{Privacy Considerations} 
& \text{I was satisfied with the privacy issues which were} \\
\textbf{} & \text{} 
& \text{considered in the procedure of the experiments.} \\
\midrule

\textbf{$Theme_9$} & \text{Self-Emotions Awareness} 
& \text{The results brought a sense of self-awareness about my} \\
\textbf{} & \text{} 
& \text{emotions in my tweets!} \\
\midrule

\textbf{$Theme_{10}$} & \text{Characteristic Type Personalization} 
& \text{I think one other thing that should be taken into account} \\
\textbf{} & \text{} 
& \text{is that consider whether the individual is introverted} \\
\textbf{} & \text{} 
& \text{or extroverted.} \\
\bottomrule

\end{tabular}
\label{tab:qualitative}
\end{table*}

\section{Discussion}

This research is motivated by insights from previous research, as highlighted in Section \ref{sec:intro}, emphasizing the importance of emotions as a fundamental component of individuals' personalities. Recognizing that emotions are not expressed in the same way across all contexts and considering the pivotal role of social media in people's lives, the goal was to quantify the potential divergence between individuals' emotions in real life and on social media platforms. It is noteworthy that despite the absence of comparable studies addressing a similar research question, we developed a framework and, through post-evaluations, assessed the effectiveness of our pipeline. The prior works \cite{kuvsen2017identifying, gaind2019emotion} are the most similar papers to ours, with two main differences in research question, alongside the differences in contributions:
\begin{itemize}
    \item We have involved human in our proposed pipeline and it is more human-centered compared to the pipelines suggested in these two works.
    \item We want to compare two different personalities in the real world and on the social media, but these papers only intend to detect and analyze emotions on social media.
\end{itemize}
Now, in the following section, our contributions, findings, and limitations of this study are discussed, which could be addressed in future work.

\subsection{Theoretical and Practical Contributions}

We made several contributions in theory and practice. Firstly, a novel dataset of Persian tweets is collected and labeled meticulously by humans in 5 emotion classes. Additionally, a hybrid model is designed, combining a Transformers-based approach for Persian sentiment analysis with a rule-based module to discern the sentiments expressed in tweets, achieving an accuracy of 74.08\% on the test set for a five-class classification problem. Moreover, leveraging the sentiment analysis module and cutting-edge models for facial expression analysis alongside human perspectives, an innovative pipeline was introduced. This pipeline aimed to quantify the extent to which individuals exhibit similar emotions in the real world compared to their expressions on social media. To validate these findings, statistical tests were conducted, and the differences in sentiment proportions were found to be statistically significant. Furthermore, using the proposed pipeline, an experiment was conducted, encompassing N=105 participants and 393 of their acquaintances (including friends, family members, partners, etc.).  Throughout the experiment, over 5000 tweets and 2000 images were collected to measure individuals' emotions on social media. Finally, a post-experiment assessment, employing both qualitative and quantitative methods, was conducted to evaluate the study. The results indicated a satisfaction rate of approximately 93\% among the participants, with various shared insights on the limitations of this study which facilitate shaping the direction for future studies.

\subsection{Findings}
Our findings provide valuable insights into the emotional discrepancies between real-life and online expressions.

\begin{itemize}
    \item The innovative pipeline we developed exposes substantial gaps between individuals' real-world emotions and their online personas. This approach has proven both effective and adaptable, suggesting its potential for widespread application across various platforms and languages. By extending this methodology to different contexts, we could gain a more nuanced understanding of how emotions are expressed and perceived in diverse digital environments.

    \item The user study conducted to address RQ \ref{RQ} reveals a significant divergence between emotions conveyed through images and those expressed in tweets. While tweets generally reflect real-life emotions with a high degree of accuracy, images often fail to capture individuals' true feelings. This indicates that textual content may be a more reliable indicator of emotional states than visual content. The statistical significance of these differences highlights the need to carefully consider the medium when analyzing emotional expression. These results advocate for further research into how various formats and platforms affect emotional representation and suggest that our understanding of online behavior should account for the distinct ways in which different media convey emotions.

    \item Additionally, the significant statistical differences observed in sentiment distributions between images, tweets, and friends' perceptions imply that each medium and perspective offers unique insights into emotional expression. These variations emphasize the need for multi-faceted approaches in emotional research, combining various sources of data to achieve a holistic view of how emotions are communicated across different contexts.
\end{itemize}

\subsection{Limitations and Future Work}

Besides our contributions to answering the primary research question, several limitations impact the findings and their validity:

\begin{itemize}
    \item One key limitation of our work is its focus on a specific language and region for the experiment. This limitation may restrict the generalizability of our findings, as emotional expressions can vary significantly across different languages and cultures. Future research should extend the core framework to include multiple languages and regions, employing analogous sentiment analysis modules. Conducting international studies could provide insights into how emotional expression differs across cultures and regions, thereby enhancing the robustness and applicability of our results.

    \item Another limitation is the scope of multimedia types considered. Our study focused primarily on images and text, while other forms of multimedia, such as GIFs, memes, and videos, were not included. These additional types of content might offer unique insights into emotional expression and could influence the overall validity of our conclusions. Expanding the research to include a broader range of multimedia types could address this gap and provide a more comprehensive understanding of emotional representation.

    \item The generalizability of our findings to different social media platforms was limited, as the study concentrated on a single platform X. This focus may overlook platform-specific factors that influence emotional expression. Future work should explore various social media platforms to understand how different digital environments affect emotional representation. Such studies would enhance the scope and applicability of our findings, offering a more nuanced view of emotional expression across diverse online contexts.

    \item The study's reliance on predefined sentiment labels and models may restrict the diversity of results. Different labeling schemes or sentiment analysis models might yield varying insights. Future research should explore alternative labeling approaches and models to validate and enrich the findings. Testing new methodologies could provide different perspectives on emotional analysis and contribute to a more thorough understanding of emotional expressions.

    \item Participant age could also impact the study's findings. Our sample may not represent the full spectrum of age groups, which could skew the results. To address this, future studies should include a broader range of age demographics to ensure that findings are representative of different age groups and their unique emotional expression patterns.

    \item A potential area for enhancement in our work is the integration of more recent state-of-the-art models and the expansion of our evaluation to include a wider variety of architectures. As the field of Natural Language Processing evolves rapidly, with new models and techniques emerging frequently, it is natural to anticipate that future advancements may offer improvements in performance and efficiency. Nevertheless, our current approach yields strong results—supported by participant feedback—and provides a solid foundation within the scope of the tools and resources available at the time of the study. Incorporating newer models in future work may further enrich the analysis and strengthen comparative benchmarks.

    \item Lastly, our study did not account for the temporal aspects of social media content or individual differences in emotional expression. Incorporating time-limited content and personalizing the experiment based on individuals' emotional types could provide more tailored insights. Future research focusing on these elements could refine the analysis and improve the understanding of how emotions evolve over time and across different personal contexts.
\end{itemize}

\section{Conclusion}

In this study, recognizing the significance of people's emotions as an important aspect of their personality and acknowledging the potential disparity in emotional expression across various situations, including both real and virtual lives, an innovative framework was introduced. This framework aims to measure the alignment between individuals' emotions on social media and in their real lives. The proposed pipeline integrates human perspectives with AI-based modules to offer a comprehensive approach. Consequently, utilizing the proposed pipeline, a user study was conducted involving N=105 participants. The study measured emotional alignment in two distinct environments of their lives. The post-experiment analysis revealed a 93\% satisfaction rate with our methodology, validating its effectiveness. Statistical analyses, including independent two-sample t-tests, confirmed the significance of the differences observed in sentiment proportions across images, tweets, and friends' perceptions, reinforcing the robustness of our findings. However, the qualitative analysis offered valuable insights that can help guide and inspire future research in this area.

\bibliographystyle{IEEEtran}
\bibliography{references}

 





\newpage
\appendix

\section{Details of the Distance Metrics}
\label{appendix:distance_metrics}
\subsection{KL-Divergence}
KL-Divergence (Kullback-Leibler Divergence) \cite{hershey2007approximating} measures the distance between two probability distributions by summing the divergence of each distribution from the other. It is commonly used but does not handle small changes effectively and lacks boundedness, making it less suitable for our application. The formula used for calculation is:

\[
D_{KL}(P \parallel Q) = \sum_{i} P(i) \log \left( \frac{P(i)}{Q(i)} \right)
\tag{2}
\]

KL-Divergence can be problematic because it tends to give unbounded values, meaning that a small difference between two similar distributions can result in a large distance, making it hard to interpret. For example, comparing distributions \( P = [0.4, 0.1, 0.1, 0.2, 0.2] \) and \( Q = [0.3, 0.2, 0.2, 0.2, 0.1] \) yields a result of approximately 0.57, which fails to provide a meaningful measure of similarity for distributions that are closely related.

\subsection{Jensen-Shannon Divergence}
Jensen-Shannon Divergence \cite{menendez1997jensen} is a symmetrized and smoothed version of KL-Divergence. It improves upon KL-Divergence by being bounded, but it still does not capture small variations as effectively as Earth Mover's Distance. The formula used for calculation is:

\[
D_{JS}(P \parallel Q) = \frac{1}{2} \left[ D_{KL}(P \parallel M) + D_{KL}(Q \parallel M) \right]
\tag{3}
\]

where \( M = \frac{1}{2}(P + Q) \). While Jensen-Shannon Divergence ensures that the distance is bounded and symmetrized, it does not sufficiently capture small differences between distributions. In our example of comparing \( P = [0.4, 0.1, 0.1, 0.2, 0.2] \) and \( Q = [0.3, 0.2, 0.2, 0.2, 0.1] \), the result was about 0.06, which is not sufficiently sensitive to minor differences between the distributions. This makes it less effective for fine-grained comparisons in our study.

These two metrics, while useful in other contexts, did not meet the criteria for our analysis, as they either lacked boundedness or failed to capture small variations effectively. In contrast, using the same example, \textbf{Earth Mover’s Distance} yielded a result of 0.4. This value reflects a bounded and sensitive measure of dissimilarity, making it the most appropriate metric for assessing the similarity between sentiment distributions.

\section{Statistical Test Datasets}
\label{appendix:sentiment_analysis}

{\small

\begin{adjustwidth}{-1cm}{-1cm} 
\onecolumn
\begin{longtable}{|c|c|c|c|c|c|c|c|c|c|c|c|c|c|c|c|c|c|c|}
\caption{Sentiment Analysis Results on Real World, Tweets, and Images of all the Participants}
\\

\hline
\textbf{A\footnote{A: Participant Number: Identifier for each individual in the study.}} & 
\textbf{B\footnote{B: Images Portion of Happy: Percentage of images categorized as 'Happy'.}} & 
\textbf{C\footnote{C: Images Portion of Sad: Percentage of images categorized as 'Sad'.}} & 
\textbf{D\footnote{D: Images Portion of Angry: Percentage of images categorized as 'Angry'.}} & 
\textbf{E\footnote{E: Images Portion of Neutral: Percentage of images categorized as 'Neutral'.}} & 
\textbf{F\footnote{F: Images Portion of Intense Emotions: Percentage of images categorized as 'Intense Emotions'.}} & 
\textbf{G\footnote{G: Tweets Portion of Happy: Percentage of tweets categorized as 'Happy'.}} & 
\textbf{H\footnote{H: Tweets Portion of Sad: Percentage of tweets categorized as 'Sad'.}} & 
\textbf{I\footnote{I: Tweets Portion of Angry: Percentage of tweets categorized as 'Angry'.}} & 
\textbf{J\footnote{J: Tweets Portion of Neutral: Percentage of tweets categorized as 'Neutral'.}} & 
\textbf{K\footnote{K: Tweets Portion of Intense Emotions: Percentage of tweets categorized as 'Intense Emotions'.}} & 
\textbf{L\footnote{L: Real World Portion of Happy: Percentage of real-world data categorized as 'Happy'.}} & 
\textbf{M\footnote{M: Real World Portion of Sad: Percentage of real-world data categorized as 'Sad'.}} & 
\textbf{N\footnote{N: Real World Portion of Angry: Percentage of real-world data categorized as 'Angry'.}} & 
\textbf{O\footnote{O: Real World Portion of Neutral: Percentage of real-world data categorized as 'Neutral'.}} & 
\textbf{P\footnote{P: Real World Portion of Intense Emotions: Percentage of real-world data categorized as 'Intense Emotions'.}} & 
\textbf{Q\footnote{Q: Distance Between Images and Real World (Earth Mover’s Distance): Measure of dissimilarity between image sentiment distribution and real-world sentiment distribution.}} & 
\textbf{R\footnote{R: Distance Between Tweets and Real World (Earth Mover’s Distance): Measure of dissimilarity between tweet sentiment distribution and real-world sentiment distribution.}} & 
\textbf{S\footnote{S: Distance Between Images and Tweets (Earth Mover’s Distance): Measure of dissimilarity between image sentiment distribution and tweet sentiment distribution.}} \\

\hline
P1 & 0.72 & 0.03 & 0.07 & 0.0 & 0.18 & 0.12 & 0.18 & 0.14 & 0.3 & 0.26 & 0.38 & 0.11 & 0.15 & 0.1 & 0.26 & 0.3 & 0.84 & 0.15 \\ 
P2 & 0.37 & 0.03 & 0.02 & 0.13 & 0.45 & 0.2 & 0.36 & 0.18 & 0.14 & 0.12 & 0.18 & 0.17 & 0.21 & 0.22 & 0.22 & 0.24 & 0.72 & 0.48 \\ 
P3 & 0.71 & 0.07 & 0.0 & 0.0 & 0.22 & 0.04 & 0.24 & 0.24 & 0.31 & 0.16 & 0.17 & 0.23 & 0.2 & 0.23 & 0.16 & 0.04 & 0.73 & 0.2 \\ 
P4 & 0.95 & 0.05 & 0.0 & 0.0 & 0.0 & 0.14 & 0.3 & 0.14 & 0.26 & 0.16 & 0.17 & 0.25 & 0.16 & 0.24 & 0.18 & -0.39 & 0.86 & -0.3 \\ 
P5 & 0.78 & 0.0 & 0.01 & 0.0 & 0.21 & 0.18 & 0.16 & 0.24 & 0.22 & 0.2 & 0.28 & 0.15 & 0.17 & 0.24 & 0.17 & 0.0 & 0.89 & -0.08 \\ 
P6 & 0.25 & 0.27 & 0.1 & 0.22 & 0.16 & 0.1 & 0.08 & 0.2 & 0.28 & 0.34 & 0.32 & 0.12 & 0.12 & 0.17 & 0.28 & 0.81 & 0.88 & 0.8 \\ 
P7 & 0.63 & 0.0 & 0.0 & 0.0 & 0.37 & 0.24 & 0.24 & 0.08 & 0.28 & 0.16 & 0.29 & 0.18 & 0.14 & 0.16 & 0.23 & 0.03 & 0.86 & 0.04 \\ 
P8 & 0.82 & 0.03 & 0.01 & 0.03 & 0.11 & 0.16 & 0.2 & 0.18 & 0.22 & 0.24 & 0.29 & 0.12 & 0.13 & 0.18 & 0.28 & -0.06 & 0.78 & -0.15 \\ 
P9 & 0.5 & 0.08 & 0.05 & 0.07 & 0.3 & 0.14 & 0.38 & 0.14 & 0.1 & 0.24 & 0.19 & 0.25 & 0.12 & 0.22 & 0.22 & 0.34 & 0.7 & 0.64 \\ 
P10 & 0.26 & 0.01 & 0.01 & 0.0 & 0.72 & 0.04 & 0.26 & 0.38 & 0.18 & 0.14 & 0.16 & 0.29 & 0.19 & 0.16 & 0.2 & 0.02 & 0.7 & 0.31 \\ 
P11 & 0.5 & 0.09 & 0.02 & 0.08 & 0.31 & 0.28 & 0.36 & 0.12 & 0.14 & 0.1 & 0.2 & 0.2 & 0.1 & 0.28 & 0.22 & 0.37 & 0.72 & 0.65 \\ 
P12 & 0.28 & 0.03 & 0.01 & 0.0 & 0.68 & 0.16 & 0.24 & 0.1 & 0.26 & 0.24 & 0.3 & 0.2 & 0.06 & 0.2 & 0.23 & 0.15 & 0.82 & 0.07 \\ 
P13 & 0.36 & 0.13 & 0.13 & 0.0 & 0.38 & 0.09 & 0.09 & 0.29 & 0.4 & 0.14 & 0.21 & 0.18 & 0.16 & 0.25 & 0.2 & 0.43 & 0.55 & 0.77 \\ 
P14 & 0.58 & 0.13 & 0.03 & 0.0 & 0.26 & 0.16 & 0.16 & 0.14 & 0.4 & 0.14 & 0.23 & 0.13 & 0.2 & 0.25 & 0.2 & 0.27 & 0.67 & 0.44 \\ 
P15 & 0.76 & 0.01 & 0.02 & 0.01 & 0.2 & 0.2 & 0.2 & 0.22 & 0.24 & 0.14 & 0.21 & 0.18 & 0.14 & 0.22 & 0.24 & -0.03 & 0.96 & -0.04 \\ 
P16 & 0.22 & 0.15 & 0.12 & 0.01 & 0.5 & 0.14 & 0.12 & 0.3 & 0.34 & 0.1 & 0.2 & 0.15 & 0.13 & 0.33 & 0.19 & 0.62 & 0.78 & 0.66 \\ 
P17 & 0.86 & 0.03 & 0.0 & 0.0 & 0.11 & 0.16 & 0.22 & 0.1 & 0.28 & 0.24 & 0.12 & 0.31 & 0.18 & 0.18 & 0.2 & -0.09 & 0.84 & -0.15 \\ 
P18 & 0.69 & 0.03 & 0.01 & 0.0 & 0.27 & 0.08 & 0.16 & 0.12 & 0.4 & 0.24 & 0.29 & 0.11 & 0.09 & 0.23 & 0.28 & 0.2 & 0.75 & 0.36 \\ 
P19 & 0.22 & 0.26 & 0.12 & 0.1 & 0.31 & 0.14 & 0.08 & 0.28 & 0.3 & 0.2 & 0.15 & 0.22 & 0.24 & 0.18 & 0.21 & 0.77 & 0.76 & 0.91 \\ 
P20 & 0.46 & 0.07 & 0.06 & 0.01 & 0.4 & 0.1 & 0.26 & 0.26 & 0.16 & 0.22 & 0.22 & 0.21 & 0.16 & 0.16 & 0.24 & 0.2 & 0.87 & 0.31 \\ 
P21 & 0.56 & 0.03 & 0.04 & 0.01 & 0.36 & 0.2 & 0.2 & 0.2 & 0.14 & 0.26 & 0.25 & 0.13 & 0.13 & 0.25 & 0.25 & 0.15 & 0.8 & 0.07 \\ 
P22 & 0.64 & 0.03 & 0.01 & 0.0 & 0.32 & 0.23 & 0.2 & 0.14 & 0.11 & 0.31 & 0.22 & 0.19 & 0.15 & 0.18 & 0.27 & 0.05 & 0.85 & 0.17 \\ 
P23 & 0.64 & 0.0 & 0.0 & 0.0 & 0.36 & 0.14 & 0.26 & 0.18 & 0.2 & 0.22 & 0.22 & 0.16 & 0.1 & 0.25 & 0.26 & 0.03 & 0.88 & -0.04 \\ 
P24 & 0.45 & 0.06 & 0.05 & 0.01 & 0.42 & 0.06 & 0.22 & 0.32 & 0.12 & 0.28 & 0.28 & 0.16 & 0.18 & 0.18 & 0.21 & 0.21 & 0.68 & 0.44 \\ 
P25 & 0.44 & 0.07 & 0.05 & 0.1 & 0.34 & 0.17 & 0.17 & 0.08 & 0.25 & 0.33 & 0.25 & 0.17 & 0.15 & 0.18 & 0.25 & 0.43 & 0.84 & 0.6 \\ 
P26 & 0.18 & 0.01 & 0.01 & 0.0 & 0.8 & 0.23 & 0.23 & 0.14 & 0.34 & 0.06 & 0.3 & 0.2 & 0.16 & 0.12 & 0.22 & 0.0 & 0.84 & 0.07 \\ 
P27 & 0.94 & 0.0 & 0.0 & 0.0 & 0.06 & 0.15 & 0.32 & 0.09 & 0.35 & 0.09 & 0.29 & 0.14 & 0.19 & 0.19 & 0.2 & -0.3 & 0.63 & -0.17 \\ 
P28 & 0.39 & 0.05 & 0.06 & 0.03 & 0.48 & 0.28 & 0.16 & 0.24 & 0.2 & 0.12 & 0.0 & 0.0 & 0.0 & 0.0 & 0.0 & 0.0 & 0.0 & 0.31 \\ 
P29 & 0.8 & 0.09 & 0.03 & 0.0 & 0.08 & 0.16 & 0.22 & 0.24 & 0.16 & 0.22 & 0.23 & 0.14 & 0.23 & 0.11 & 0.28 & -0.03 & 0.86 & -0.12 \\ 
P30 & 0.51 & 0.08 & 0.08 & 0.02 & 0.31 & 0.28 & 0.18 & 0.18 & 0.12 & 0.24 & 0.27 & 0.16 & 0.1 & 0.19 & 0.29 & 0.46 & 0.91 & 0.4 \\ 
P31 & 0.38 & 0.45 & 0.0 & 0.0 & 0.18 & 0.28 & 0.3 & 0.08 & 0.18 & 0.16 & 0.23 & 0.24 & 0.14 & 0.09 & 0.3 & 0.43 & 0.88 & 0.51 \\ 
P32 & 0.46 & 0.09 & 0.05 & 0.04 & 0.36 & 0.17 & 0.23 & 0.15 & 0.25 & 0.21 & 0.25 & 0.18 & 0.13 & 0.21 & 0.23 & 0.32 & 0.96 & 0.31 \\ 
P33 & 0.36 & 0.1 & 0.13 & 0.12 & 0.29 & 0.24 & 0.26 & 0.1 & 0.26 & 0.14 & 0.32 & 0.12 & 0.16 & 0.13 & 0.27 & 0.88 & 0.82 & 0.74 \\ 
P34 & 0.37 & 0.17 & 0.09 & 0.11 & 0.26 & 0.08 & 0.2 & 0.18 & 0.16 & 0.38 & 0.23 & 0.29 & 0.14 & 0.19 & 0.16 & 0.76 & 0.81 & 0.86 \\ 
P35 & 0.57 & 0.01 & 0.01 & 0.0 & 0.41 & 0.14 & 0.16 & 0.16 & 0.42 & 0.12 & 0.18 & 0.17 & 0.22 & 0.24 & 0.19 & -0.03 & 0.64 & 0.2 \\ 
P36 & 0.32 & 0.08 & 0.01 & 0.0 & 0.59 & 0.14 & 0.26 & 0.12 & 0.28 & 0.2 & 0.15 & 0.32 & 0.09 & 0.2 & 0.23 & 0.29 & 0.88 & 0.26 \\ 
P37 & 0.23 & 0.01 & 0.01 & 0.03 & 0.72 & 0.22 & 0.12 & 0.2 & 0.36 & 0.1 & 0.26 & 0.17 & 0.17 & 0.24 & 0.17 & 0.07 & 0.73 & 0.26 \\ 
P38 & 0.46 & 0.05 & 0.08 & 0.08 & 0.33 & 0.08 & 0.3 & 0.04 & 0.1 & 0.48 & 0.15 & 0.27 & 0.12 & 0.2 & 0.25 & 0.47 & 0.49 & 0.92 \\ 
P39 & 0.17 & 0.05 & 0.07 & 0.17 & 0.54 & 0.22 & 0.24 & 0.18 & 0.24 & 0.12 & 0.23 & 0.15 & 0.22 & 0.23 & 0.17 & 0.38 & 0.94 & 0.39 \\ 
P40 & 0.36 & 0.12 & 0.04 & 0.02 & 0.47 & 0.18 & 0.2 & 0.16 & 0.18 & 0.28 & 0.24 & 0.17 & 0.11 & 0.28 & 0.21 & 0.38 & 0.86 & 0.31 \\ 
P41 & 0.42 & 0.19 & 0.02 & 0.0 & 0.36 & 0.24 & 0.24 & 0.2 & 0.22 & 0.1 & 0.39 & 0.1 & 0.12 & 0.12 & 0.27 & 0.6 & 0.63 & 0.38 \\ 
P42 & 0.83 & 0.03 & 0.0 & 0.0 & 0.14 & 0.08 & 0.08 & 0.18 & 0.5 & 0.16 & 0.36 & 0.11 & 0.1 & 0.1 & 0.34 & 0.05 & 0.61 & 0.34 \\ 
P43 & 0.48 & 0.01 & 0.02 & 0.04 & 0.46 & 0.1 & 0.36 & 0.12 & 0.16 & 0.26 & 0.19 & 0.24 & 0.15 & 0.26 & 0.17 & 0.12 & 0.75 & 0.37 \\ 
P44 & 0.36 & 0.05 & 0.07 & 0.05 & 0.47 & 0.42 & 0.1 & 0.04 & 0.28 & 0.16 & 0.35 & 0.08 & 0.06 & 0.37 & 0.14 & 0.78 & 0.82 & 0.72 \\ 
P45 & 0.66 & 0.01 & 0.0 & 0.0 & 0.33 & 0.06 & 0.18 & 0.06 & 0.5 & 0.2 & 0.29 & 0.12 & 0.07 & 0.31 & 0.2 & 0.23 & 0.62 & 0.41 \\ 
P46 & 0.44 & 0.25 & 0.01 & 0.0 & 0.31 & 0.16 & 0.26 & 0.08 & 0.34 & 0.16 & 0.27 & 0.08 & 0.14 & 0.17 & 0.34 & 0.57 & 0.96 & 0.53 \\ 
P47 & 0.9 & 0.03 & 0.01 & 0.0 & 0.06 & 0.12 & 0.3 & 0.2 & 0.28 & 0.1 & 0.13 & 0.31 & 0.2 & 0.2 & 0.17 & -0.18 & 0.83 & -0.2 \\ 
P48 & 0.31 & 0.01 & 0.01 & 0.0 & 0.67 & 0.1 & 0.36 & 0.1 & 0.22 & 0.22 & 0.23 & 0.21 & 0.1 & 0.33 & 0.13 & 0.16 & 0.92 & 0.2 \\ 
P49 & 0.36 & 0.2 & 0.06 & 0.1 & 0.27 & 0.18 & 0.08 & 0.08 & 0.36 & 0.3 & 0.32 & 0.13 & 0.11 & 0.15 & 0.3 & 0.79 & 0.84 & 0.9 \\ 
P50 & 0.64 & 0.06 & 0.08 & 0.11 & 0.11 & 0.16 & 0.34 & 0.16 & 0.3 & 0.04 & 0.26 & 0.17 & 0.16 & 0.14 & 0.27 & 0.26 & 0.78 & 0.36 \\ 
P51 & 0.34 & 0.19 & 0.08 & 0.0 & 0.4 & 0.18 & 0.2 & 0.24 & 0.24 & 0.14 & 0.23 & 0.18 & 0.11 & 0.25 & 0.23 & 0.49 & 0.92 & 0.49 \\ 
P52 & 0.53 & 0.03 & 0.06 & 0.02 & 0.36 & 0.26 & 0.1 & 0.08 & 0.34 & 0.22 & 0.31 & 0.13 & 0.07 & 0.28 & 0.22 & 0.38 & 0.9 & 0.42 \\ 
P53 & 0.52 & 0.04 & 0.08 & 0.1 & 0.26 & 0.24 & 0.08 & 0.16 & 0.38 & 0.14 & 0.21 & 0.15 & 0.21 & 0.26 & 0.18 & 0.37 & 0.69 & 0.68 \\ 
P54 & 0.39 & 0.19 & 0.02 & 0.06 & 0.34 & 0.1 & 0.08 & 0.1 & 0.42 & 0.3 & 0.31 & 0.15 & 0.14 & 0.17 & 0.23 & 0.57 & 0.64 & 0.74 \\ 
P55 & 0.58 & 0.06 & 0.03 & 0.02 & 0.31 & 0.24 & 0.12 & 0.18 & 0.22 & 0.24 & 0.37 & 0.1 & 0.09 & 0.15 & 0.28 & 0.53 & 0.64 & 0.17 \\ 
P56 & 0.33 & 0.15 & 0.08 & 0.03 & 0.4 & 0.24 & 0.14 & 0.14 & 0.14 & 0.34 & 0.14 & 0.25 & 0.17 & 0.19 & 0.25 & 0.52 & 0.82 & 0.66 \\ 
P57 & 0.46 & 0.04 & 0.02 & 0.0 & 0.48 & 0.14 & 0.26 & 0.26 & 0.21 & 0.12 & 0.27 & 0.17 & 0.13 & 0.16 & 0.28 & 0.2 & 0.91 & 0.17 \\ 
P58 & 0.05 & 0.17 & 0.02 & 0.06 & 0.7 & 0.14 & 0.28 & 0.16 & 0.14 & 0.28 & 0.24 & 0.24 & 0.05 & 0.21 & 0.25 & 0.1 & 0.69 & 0.16 \\ 
P59 & 0.74 & 0.04 & 0.01 & 0.01 & 0.2 & 0.14 & 0.34 & 0.08 & 0.06 & 0.38 & 0.31 & 0.12 & 0.1 & 0.16 & 0.31 & 0.14 & 0.8 & 0.28 \\ 
P60 & 0.41 & 0.12 & 0.09 & 0.05 & 0.33 & 0.24 & 0.28 & 0.18 & 0.14 & 0.16 & 0.22 & 0.24 & 0.15 & 0.17 & 0.22 & 0.43 & 0.88 & 0.56 \\ 
P61 & 0.23 & 0.01 & 0.01 & 0.0 & 0.75 & 0.2 & 0.26 & 0.22 & 0.14 & 0.18 & 0.16 & 0.16 & 0.19 & 0.2 & 0.28 & 0.0 & 0.91 & 0.0 \\ 
P62 & 0.08 & 0.24 & 0.19 & 0.3 & 0.19 & 0.04 & 0.29 & 0.36 & 0.18 & 0.14 & 0.21 & 0.21 & 0.18 & 0.21 & 0.19 & 0.76 & 0.55 & 0.79 \\ 
P63 & 0.43 & 0.19 & 0.16 & 0.07 & 0.15 & 0.1 & 0.34 & 0.08 & 0.24 & 0.24 & 0.2 & 0.17 & 0.19 & 0.19 & 0.26 & 0.65 & 0.64 & 0.72 \\ 
P64 & 1.0 & 0.0 & 0.0 & 0.0 & 0.0 & 0.14 & 0.09 & 0.06 & 0.49 & 0.23 & 0.28 & 0.15 & 0.07 & 0.25 & 0.26 & -0.44 & 0.58 & -0.02 \\ 
P65 & 0.71 & 0.07 & 0.02 & 0.02 & 0.18 & 0.16 & 0.2 & 0.12 & 0.32 & 0.2 & 0.29 & 0.09 & 0.11 & 0.17 & 0.34 & 0.26 & 0.78 & 0.21 \\ 
P66 & 0.37 & 0.03 & 0.03 & 0.01 & 0.56 & 0.25 & 0.2 & 0.05 & 0.15 & 0.35 & 0.24 & 0.18 & 0.06 & 0.28 & 0.24 & 0.17 & 0.84 & 0.33 \\ 
P67 & 0.99 & 0.01 & 0.0 & 0.0 & 0.0 & 0.32 & 0.16 & 0.18 & 0.26 & 0.08 & 0.29 & 0.28 & 0.1 & 0.12 & 0.21 & -0.39 & 0.85 & -0.34 \\ 
P68 & 0.15 & 0.06 & 0.04 & 0.08 & 0.67 & 0.1 & 0.06 & 0.28 & 0.3 & 0.26 & 0.28 & 0.16 & 0.07 & 0.24 & 0.25 & 0.21 & 0.86 & 0.25 \\ 
P69 & 0.28 & 0.03 & 0.01 & 0.0 & 0.68 & 0.08 & 0.24 & 0.14 & 0.3 & 0.24 & 0.34 & 0.17 & 0.05 & 0.18 & 0.25 & 0.27 & 0.82 & 0.15 \\ 
P70 & 0.56 & 0.31 & 0.02 & 0.02 & 0.09 & 0.4 & 0.12 & 0.16 & 0.2 & 0.12 & 0.27 & 0.16 & 0.14 & 0.15 & 0.29 & 0.36 & 0.77 & 0.46 \\ 
P71 & 0.34 & 0.2 & 0.13 & 0.09 & 0.25 & 0.06 & 0.12 & 0.06 & 0.51 & 0.25 & 0.26 & 0.23 & 0.1 & 0.17 & 0.23 & 0.82 & 0.46 & 0.64 \\ 
P72 & 0.13 & 0.02 & 0.05 & 0.03 & 0.77 & 0.12 & 0.32 & 0.14 & 0.22 & 0.2 & 0.21 & 0.34 & 0.08 & 0.08 & 0.29 & 0.13 & 0.8 & 0.09 \\ 
P73 & 0.32 & 0.12 & 0.08 & 0.1 & 0.38 & 0.26 & 0.18 & 0.22 & 0.24 & 0.1 & 0.25 & 0.13 & 0.13 & 0.22 & 0.27 & 0.64 & 0.9 & 0.6 \\ 
P74 & 0.58 & 0.01 & 0.02 & 0.0 & 0.39 & 0.05 & 0.41 & 0.05 & 0.14 & 0.36 & 0.16 & 0.2 & 0.17 & 0.17 & 0.3 & 0.06 & 0.47 & 0.58 \\ 
P75 & 0.34 & 0.19 & 0.05 & 0.1 & 0.33 & 0.22 & 0.22 & 0.1 & 0.12 & 0.34 & 0.19 & 0.24 & 0.14 & 0.18 & 0.25 & 0.65 & 0.76 & 0.78 \\ 
P76 & 0.54 & 0.06 & 0.04 & 0.02 & 0.34 & 0.06 & 0.52 & 0.2 & 0.04 & 0.18 & 0.24 & 0.23 & 0.14 & 0.12 & 0.26 & 0.25 & 0.48 & 0.68 \\ 
P77 & 0.17 & 0.08 & 0.17 & 0.15 & 0.43 & 0.2 & 0.3 & 0.1 & 0.24 & 0.16 & 0.23 & 0.19 & 0.17 & 0.24 & 0.17 & 0.62 & 0.84 & 0.74 \\ 
P78 & 0.08 & 0.12 & 0.11 & 0.01 & 0.68 & 0.5 & 0.1 & 0.08 & 0.2 & 0.12 & 0.33 & 0.1 & 0.07 & 0.19 & 0.31 & 0.3 & 0.64 & 0.63 \\ 
P79 & 0.48 & 0.1 & 0.02 & 0.02 & 0.38 & 0.1 & 0.32 & 0.12 & 0.32 & 0.14 & 0.17 & 0.27 & 0.15 & 0.25 & 0.17 & 0.3 & 0.74 & 0.56 \\ 
P80 & 0.42 & 0.14 & 0.06 & 0.04 & 0.33 & 0.14 & 0.16 & 0.22 & 0.28 & 0.2 & 0.29 & 0.19 & 0.11 & 0.16 & 0.25 & 0.56 & 0.92 & 0.48 \\ 
P81 & 0.94 & 0.0 & 0.0 & 0.0 & 0.06 & 0.1 & 0.3 & 0.1 & 0.24 & 0.26 & 0.24 & 0.19 & 0.17 & 0.22 & 0.17 & -0.39 & 0.71 & -0.28 \\ 
P82 & 0.47 & 0.1 & 0.14 & 0.02 & 0.28 & 0.1 & 0.4 & 0.16 & 0.12 & 0.22 & 0.21 & 0.17 & 0.14 & 0.26 & 0.22 & 0.47 & 0.72 & 0.75 \\ 
P83 & 0.43 & 0.03 & 0.02 & 0.01 & 0.51 & 0.26 & 0.24 & 0.1 & 0.14 & 0.26 & 0.28 & 0.17 & 0.14 & 0.18 & 0.23 & 0.14 & 0.82 & 0.15 \\ 
P84 & 0.49 & 0.05 & 0.04 & 0.0 & 0.42 & 0.1 & 0.34 & 0.14 & 0.28 & 0.14 & 0.18 & 0.32 & 0.23 & 0.14 & 0.14 & 0.26 & 0.84 & 0.42 \\ 
P85 & 0.37 & 0.06 & 0.01 & 0.01 & 0.55 & 0.06 & 0.44 & 0.1 & 0.16 & 0.24 & 0.25 & 0.17 & 0.17 & 0.18 & 0.22 & 0.1 & 0.58 & 0.52 \\ 
P86 & 0.57 & 0.07 & 0.02 & 0.01 & 0.33 & 0.14 & 0.1 & 0.12 & 0.32 & 0.32 & 0.29 & 0.1 & 0.05 & 0.29 & 0.26 & 0.37 & 0.75 & 0.48 \\ 
P87 & 0.52 & 0.09 & 0.04 & 0.03 & 0.32 & 0.08 & 0.48 & 0.14 & 0.18 & 0.12 & 0.23 & 0.15 & 0.18 & 0.2 & 0.25 & 0.27 & 0.53 & 0.63 \\ 
P88 & 0.43 & 0.07 & 0.03 & 0.15 & 0.32 & 0.19 & 0.31 & 0.08 & 0.31 & 0.11 & 0.15 & 0.27 & 0.15 & 0.2 & 0.22 & 0.48 & 0.74 & 0.74 \\ 
P89 & 0.59 & 0.06 & 0.07 & 0.04 & 0.24 & 0.14 & 0.22 & 0.18 & 0.22 & 0.24 & 0.33 & 0.11 & 0.13 & 0.2 & 0.24 & 0.46 & 0.79 & 0.26 \\ 
P90 & 0.9 & 0.01 & 0.0 & 0.0 & 0.09 & 0.52 & 0.12 & 0.16 & 0.12 & 0.08 & 0.21 & 0.21 & 0.18 & 0.26 & 0.14 & -0.28 & 0.48 & 0.24 \\ 
P91 & 0.44 & 0.11 & 0.08 & 0.06 & 0.31 & 0.1 & 0.1 & 0.44 & 0.18 & 0.18 & 0.26 & 0.16 & 0.1 & 0.29 & 0.19 & 0.6 & 0.7 & 0.74 \\ 
P92 & 0.79 & 0.01 & 0.0 & 0.0 & 0.2 & 0.2 & 0.44 & 0.08 & 0.18 & 0.1 & 0.22 & 0.22 & 0.12 & 0.22 & 0.21 & -0.13 & 0.56 & 0.29 \\ 
P93 & 0.45 & 0.1 & 0.11 & 0.03 & 0.3 & 0.14 & 0.3 & 0.18 & 0.2 & 0.18 & 0.19 & 0.24 & 0.11 & 0.29 & 0.16 & 0.55 & 0.89 & 0.48 \\ 
P94 & 0.59 & 0.05 & 0.03 & 0.02 & 0.31 & 0.18 & 0.32 & 0.16 & 0.18 & 0.16 & 0.26 & 0.22 & 0.13 & 0.26 & 0.13 & 0.24 & 0.76 & 0.19 \\ 
P95 & 0.46 & 0.18 & 0.05 & 0.08 & 0.24 & 0.14 & 0.2 & 0.28 & 0.2 & 0.18 & 0.1 & 0.32 & 0.27 & 0.18 & 0.13 & 0.72 & 0.78 & 0.57 \\ 
P96 & 0.38 & 0.06 & 0.1 & 0.01 & 0.45 & 0.08 & 0.12 & 0.08 & 0.34 & 0.38 & 0.17 & 0.2 & 0.11 & 0.3 & 0.22 & 0.38 & 0.6 & 0.78 \\ 
P97 & 0.19 & 0.01 & 0.02 & 0.0 & 0.78 & 0.16 & 0.3 & 0.14 & 0.22 & 0.18 & 0.3 & 0.17 & 0.15 & 0.17 & 0.21 & 0.03 & 0.96 & 0.03 \\ 
P98 & 0.18 & 0.42 & 0.04 & 0.19 & 0.17 & 0.2 & 0.28 & 0.04 & 0.2 & 0.28 & 0.26 & 0.17 & 0.07 & 0.24 & 0.26 & 0.68 & 0.86 & 0.72 \\ 
P99 & 0.65 & 0.06 & 0.03 & 0.01 & 0.25 & 0.12 & 0.22 & 0.1 & 0.3 & 0.26 & 0.25 & 0.12 & 0.16 & 0.26 & 0.22 & 0.2 & 0.88 & 0.29 \\ 
P100 & 0.34 & 0.1 & 0.06 & 0.03 & 0.48 & 0.28 & 0.22 & 0.04 & 0.22 & 0.24 & 0.47 & 0.12 & 0.09 & 0.09 & 0.22 & 0.77 & 0.5 & 0.41 \\ 
P101 & 0.42 & 0.05 & 0.17 & 0.02 & 0.33 & 0.16 & 0.22 & 0.1 & 0.34 & 0.18 & 0.35 & 0.15 & 0.08 & 0.27 & 0.15 & 0.68 & 0.88 & 0.6 \\ 
P102 & 0.73 & 0.02 & 0.02 & 0.0 & 0.23 & 0.14 & 0.1 & 0.26 & 0.36 & 0.14 & 0.21 & 0.15 & 0.21 & 0.21 & 0.22 & -0.06 & 0.62 & 0.26 \\ 
P103 & 0.47 & 0.04 & 0.03 & 0.09 & 0.37 & 0.14 & 0.3 & 0.14 & 0.12 & 0.3 & 0.26 & 0.18 & 0.11 & 0.13 & 0.32 & 0.48 & 0.88 & 0.52 \\ 
P104 & 0.48 & 0.1 & 0.03 & 0.22 & 0.17 & 0.12 & 0.16 & 0.32 & 0.16 & 0.24 & 0.27 & 0.14 & 0.14 & 0.19 & 0.26 & 0.57 & 0.86 & 0.66 \\ 
P105 & 0.44 & 0.02 & 0.0 & 0.0 & 0.54 & 0.22 & 0.04 & 0.14 & 0.32 & 0.28 & 0.25 & 0.18 & 0.17 & 0.14 & 0.26 & 0.05 & 0.74 & 0.24 \\ 

\hline
\end{longtable}
\end{adjustwidth}
}

\vfill

\end{document}